\documentclass[tighten,amssymb,amsmath]{revtex4}
\usepackage{epsfig}
\usepackage{dcolumn}
\usepackage{bm}
\usepackage{amssymb}
\usepackage{amsmath}
\usepackage{amsfonts} 
\newcommand{\be}{\begin{equation}}
\newcommand{\ee}{\end{equation}}
\newcommand{\bea}{\begin{eqnarray}}
\newcommand{\eea}{\end{eqnarray}}
\newcommand{\ba}{\begin{array}}
\newcommand{\ea}{\end{array}}
\newcommand{\nn}{\nonumber}

\def\vev#1{\left\langle #1\right\rangle}
\catcode`\@=11
\def\lsim{\mathrel{\mathpalette\@versim<}}
\def\gsim{\mathrel{\mathpalette\@versim>}}
\def\@versim#1#2{\vcenter{\offinterlineskip
\ialign{$\m@th#1\hfil##\hfil$\crcr#2\crcr\sim\crcr } }}
\catcode`\@=12

\catcode`@=12
\begin{document}
\begin{flushright}
IFIC/10-45
\end{flushright}

\title{Neutrino phenomenology and stable dark matter with $A_4$}
\author{D.\,Meloni\footnote{davide.meloni@physik.uni-wuerzburg.de}$^{1}$, S. Morisi\footnote{morisi@ific.uv.es}$^{2}$ and E. Peinado\footnote{epeinado@ific.uv.es}$^{2}$}
\affiliation{ $^{1}$Institut f{\"u}r Theoretische Physik und Astrophysik, 
Universit{\"a}t W{\"u}rzburg, D-97074 W{\"u}rzburg.\\\\$^2$AHEP Group, Institut de F\'{\i}sica Corpuscular --
C.S.I.C./Universitat de Val{\`e}ncia 
Edificio Institutos de Paterna, Apt 22085, E--46071 Valencia, Spain }

\begin{abstract}
We present a model based on the $A_4$ non-abelian discrete symmetry leading to a predictive five-parameter neutrino mass
matrix and providing a stable dark matter candidate. We found an interesting correlation among the atmospheric
and the reactor angles which predicts  
$\theta_{23}\sim \pi/4$ for very small reactor angle and deviation from maximal atmospheric mixing for large $\theta_{13}$. 
Only normal neutrino mass spectrum is possible and the effective mass entering the neutrinoless double beta decay rate is constrained to be  $|m_{ee}|>4 \cdot 10^{-4}~\text{eV}$.

\end{abstract}

\maketitle

\section{\label{int}Introduction}

Neutrino oscillation and dark matter (DM) are so far the most important evidences
of physics beyond the standard model. 
Many models for neutrino masses and mixings have been studied in literature, some of them with the aim of reproducing the 
tri-bimaximal mixing pattern (TBM) \cite{Harrison:2002er} observed, within one standard deviation, in the recent neutrino oscillation data \cite{Schwetz:2008er,Fogli:2005cq}, namely
$\sin^2\theta_{12}=1/3$, $\sin^2\theta_{23}=1/2$ and $\sin^2\theta_{13}=0$.
In this context, non-Abelian discrete flavor symmetries  have been largely used since, especially in models
based on $A_4$ \cite{Altarelli:2010gt} and $S_4$ \cite{Bazzocchi:2008ej} permutation groups, the TBM limit can be easily reproduced.

While we have robust evidence for dark matter from rotation curves of spiral galaxies, gravitational lensing,
WMAP measurament, CMB anisotropy, structure formation, X-ray observations, bullet-clusters~~\cite{Bertone:2004pz},
we still do not have neither theoretical nor experimental indications about the nature of it; also the mechanism for DM stability is still not understood yet.
Many models assume {\it ad hoc} extra Abelian symmetries
\cite{adhoc} in order to stabilize the dark matter. Such a symmetry can arise from the
spontaneous breaking of a non-Abelian continuous symmetry, see for instance ~\cite{Joseph:2009bq}, or from the breaking of Abelian $U(1)$ group \cite{Frigerio:2009wf,Kadastik:2009dj,Batell:2010bp}
as, for example, in grand unified frameworks or in 
supersymmetry~\cite{Martin:1992mq}.

Recently a model where the stability of the DM arises from the spontaneous breakdown of a non-abelian discrete flavor symmetry 
was proposed~\cite{Hirsch:2010ru}. Such a model is based on the property that the $A_4$ group
is spontaneously  broken into its subgroup $Z_2$ which is responsably for the stability (for a paper 
with decaying dark matter with discrete non-Abelian symmetry see \cite{Esteves:2010sh,Haba:2010ag,Daikoku:2010ew}). 
In \cite{Hirsch:2010ru}
the same $Z_2$ is also acting in the neutrino sector, giving a vanishing  reactor angle $\theta_{13}=0$ and allowing only the inverted
hierarchy for the neutrino mass spectrum. 

However these properties are model-dependent features and cannot be considered as general results
of models where the  stability of the dark matter is justified by non-abelian discrete flavor symmetries. In this paper we want 
to provide an explicit example of a model, based on a simple extention of  \cite{Hirsch:2010ru},
with a richer neutrino phenomenology, predicting a normal mass ordering and  $\theta_{13}\ne 0$.

The paper is organized as follows:
in section \ref{themodel} we give the field content of our model and derive 
the neutrino mass matrix, whose phenomenological implications are discussed in section \ref{pheno}; in section \ref{conc} we draw our conclusions.

\section{The model}
\label{themodel}
In Tab.(\ref{tab1}) we summarized the model quantum numbers.
In contrast to \cite{Hirsch:2010ru}, the right-handed neutrino $N_4$ is
assigned to $1'$ instead of $1$ and we introduced one more right-handed neutrino $N_5$
assigned to $1''$. 
\begin{table}[h!]
\begin{center}
\begin{tabular}{|c|c|c|c|c|c|c|c|c|c||c|c|}
\hline
&$\,L_e\,$&$\,L_{\mu}\,$&$\,L_{\tau}\,$&$\,\,l_{e}^c\,\,$&$\,\,l_{{\mu}}^c\,\,$&$\,\,l_{{\tau}}^c\,\,$&$N_{T}\,$&$\,N_4\,$&$\,N_5\,$&$\,H\,$&$\,\eta\,$\\
\hline
$SU(2)$&2&2&2&1&1&1&1&1&1&2&2\\
\hline
$A_4$ &$1$ &$1^\prime$&$1^{\prime \prime}$&$1$&$1^{\prime \prime}$&$1^\prime$&$3$ &$1'$ &$1''$ &$1$&$3$\\
\hline
\end{tabular}\caption{\it Summary of  relevant  model quantum numbers}\label{tab1}
\end{center}
\end{table}

\noindent
The operators needed to generate neutrinos masses are the following:
\begin{eqnarray}\label{lag}
\mathcal{L}&=& y_1^\nu L_e(N_T\eta)_{1}+y_2^\nu L_\mu(N_T\eta)_{1''}+y_3^\nu L_\tau(N_T\eta)_{1'}+\nonumber\\
&&+y_4^\nu L_\tau N_4 H+y_5^\nu L_\mu N_5 H+ M_1 N_TN_T+M_2 N_4 N_5+
\mbox{h.c.}\nonumber
\end{eqnarray}
The scalar and the lepton charged current sectors of our
model are the same as in  \cite{Hirsch:2010ru} and we refer to that paper for all the details.
After electroweak symmetry breaking, two of the Higgs doublets contained in the $A_4$ triplet $\eta$ do not take
vev and we have: 
\begin{equation}
  \vev{ H^0}=v_h\ne 0,~~~~ \vev{ \eta^0_1}=v_\eta \ne 0~~~~
\vev{ \eta^0_{2,3}}=0\,.
\end{equation} 
As a consequence, the Dirac and Majorana neutrino mass matrices are:
\begin{eqnarray}
\label{Eq:mass}
 m_{D} = \left(
  \begin{array}{ccccc}
  y_1^\nu  v_\eta& 0 & 0 & 0 & 0\\
  y_2^\nu  v_\eta& 0 & 0 & 0 & y_5^\nu  v_h\\
  y_3^\nu  v_\eta& 0 & 0 & y_4^\nu  v_h& 0
  \end{array}
  \right) 
  \qquad 
  m_{M} = \left(
  \begin{array}{ccccc}
  M_1 & 0 & 0 & 0 & 0\\
  0 & M_1 & 0 & 0 & 0\\
  0 & 0 & M_1 & 0 & 0\\
  0 & 0 & 0 & 0 & M_2 \\
  0 & 0 & 0 & M_2 & 0 
  \end{array}
  \right) \,.
  \qquad 
\end{eqnarray}
Using the See-Saw type-I formula we get the following light  neutrino mass matrix:
\begin{eqnarray}
\label{Eq:mass2}
 m_{\nu} = \left(
  \begin{array}{ccc}
  a^2 & a b & a c\\
  a b  & b^2 & b c + k \\
  a c  & b c + k  & c^2
  \end{array}
  \right), 
\end{eqnarray}
where we defined:
\begin{eqnarray} 
a &=& \frac{y_1^\nu  v_\eta}{\sqrt{M_1}} \nn \qquad b = \frac{y_2^\nu  v_\eta}{\sqrt{M_1}} \nn \\  && \label{coeff} \\ 
c &=& \frac{y_3^\nu  v_\eta}{\sqrt{M_1}} \nn \qquad k = \frac{y_4^\nu y_5^\nu  v_h^2}{M_2}  \nn    \,.
\end{eqnarray} 
As discussed in more details in \cite{Hirsch:2010ru}, the group $A_4$ is broken by the vev
$\vev{ \eta^0_1}\sim (1,0,0)$ down to the subgroup
$Z_2$, generated by the diagonal $A_4$ generator $S=\text{Diag}\{1,-1,-1 \}$; the $Z_2$ 
symmetry acts on the triples fields in the following way: 
\begin{equation}\label{residualZ2}
Z_2:\quad
\begin{array}{lcrlcrlcr}
N_2 &\to& -N_2\,,\quad& h_2 &\to& -h_2\,, \quad&A_2 &\to& -A_2\,, \\   
N_3 &\to& -N_3\,,\quad& h_3 &\to& -h_3\,,\quad &A_3&\to& -A_3\,,  
\end{array}
\end{equation}
where $h_i$ and $A_i$ are respectively the CP-odd and CP-even component of the Higgs doublet 
$\eta_i$ for $i=2,3$ and $N_{2,3}$ are the component of the triplet $N_T$. 
This residual symmetry is responsible for the stability of the lightest combination of $h_2$ and $
h_3$, which is the dark matter candidate.
In fact it couples only to heavy right-handed neutrinos and not to quarks, supposed
to be singlets under $A_4$. Such a  scalar dark matter candidate is potentially detectable in nuclear recoil 
experiments \cite{Angle:2007uj,Ahmed:2009zw}.

\section{Numerical analysis of the full mass matrix} 
\label{pheno}
\begin{figure}[h!] 
\begin{center} 
\includegraphics[width=10cm]{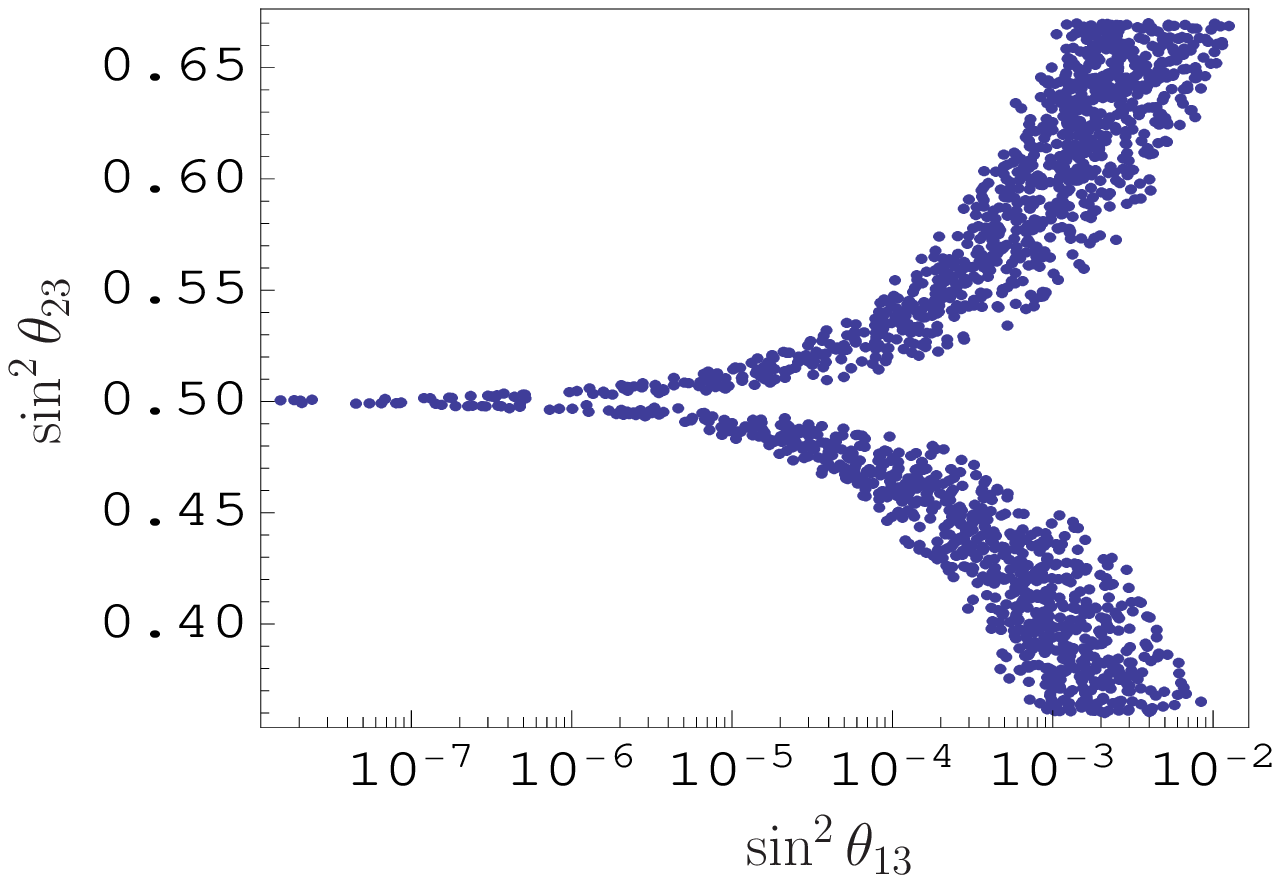}
\vspace{.2cm}
\caption{\it Correlation among the $\theta_{13}$ and $\theta_{23}$ angles as predicted by the model.}
\label{t13-th23}
\end{center}
\end{figure}

The most general complex and symmetric matrix has twelve independent real parameters which, after readsorbing the unphysical phases, reduce to nine. 
The matrix in eq.(\ref{Eq:mass2}) has four complex parameters and then five real independent parameters:  
the moduli of $a,b$ and $c$ and the modulus and phase of the combination $bc+k = d\, e^{i\, \phi_d}$.
In terms of them, the matrix in eq. (\ref{Eq:mass2}) is rewritten in the following form:
\begin{eqnarray}
\label{Eq:mass22}
 m_{\nu} = \left(
  \begin{array}{ccc}
  a^2 & a b & a c\\
  a b  & b^2 & d e^{i\phi_d}\\
  a c  & d e^{i\phi_d}  & c^2
  \end{array}
  \right). 
\end{eqnarray}
We can relate three of the five parameters in eq. (\ref{Eq:mass22})  (i.e. $a,b$ and $c$)  with the neutrino masses using the invariant quantities of the hermitian matrix $M_\nu^2=m_\nu\,m^\dagger_\nu$:
\bea
{\rm Tr} \,(M_\nu^2)  &=& T = a^4+2 a^2 \left(b^2+c^2\right)+b^4+c^4+2 d^2 =  \nn \\
&&\quad =m^2_{\nu_1}+m^2_{\nu_2}+m^2_{\nu_3} \nn \\
{\rm det}\,(M_\nu^2)&=& a^4 \left(b^2 c^2-2 b\, c\, d \,\cos \phi_d+d^2\right)^2 =  \nn \\
&&\label{invariants}\quad =m^2_{\nu_1}\,m^2_{\nu_2}\,m^2_{\nu_3} \\
\frac{1}{2}\left[T^2 - {\rm Tr} (M_\nu^2\,M_\nu^2)\right]\nn &=&
\left(b^2 c^2-2 b\, c\, d \,\cos \phi_d+d^2\right)\cdot \\ &&\nn
\left[2 a^4+2 a^2 \left(b^2+c^2\right)+b^2 c^2+2 b c d \cos \phi_d+ d^2\right]  \nn \\
&&\quad = m^2_{\nu_1} \,(m^2_{\nu_2}+m^2_{\nu_3}) + m^2_{\nu_2} m^2_{\nu_3}\, .\nn
\eea
where the symbols Tr and Det refer to the trace and determinant of a matrix, respectively.
The expressions of $a,b$ and $c$ in terms of neutrino masses, derived from the inversion of the 
previous relations, produce lengthly formulae that we do not present here but included in our numerical simulations. 
As independent parameters, we decide to take $m_{\nu_1}$, 
the two square mass differences $\Delta m^2_{sol}$ and $\Delta m^2_{atm}$ and the other two parameters of the neutrino mass matrix,
i.e. $d$ and $\phi_d$; to study the correlations among the neutrino observables predicted by the model, 
we perfom a parameter space scan extracting randomly the previous parameters in the following intervals:
\bea
&d \in [-1,1], \qquad \phi_d\in [-\pi,\pi),\qquad  m_{\nu_1} \in [10^{-5},1]\; {\rm eV}\,,& \nn \\
&&\nn\\
&\begin{array}{rcl}
7.1 \times 10^{-5}\, {\rm eV^2}<&\Delta m^2_{sol}& < 8.3 \times 10^{-5}\, {\rm eV^2}\,, \nn \\ &&  \nn  \\ 
0.027 <&r=\frac{\Delta m^2_{sol}}{|\Delta m^2_{atm}|} &< 0.040  \,. \nn 
\end{array}&
\eea
Then, we require that the obtained mixing angles are within their current 3$\sigma$ confidence level \cite{Schwetz:2008er}. 
We carefully checked that $|d|>1$ is not compatible with data. 
We made the calculation for both inverted and normal neutrino mass hierarchies finding that only the latter is allowed in our framework (we have analitically 
discussed  this point in the appendices: Appendix \ref{real} for the real case, Appendix \ref{mutau} for the 
the $\mu-\tau$ invariant case and Appendix \ref{app} for the exact TBM case). 
Among all possible correlations, we found that the $(\theta_{13}-\theta_{23})$ one is quite interesting, as shown in  fig.(\ref{t13-th23}).
As we can see, for very small $\theta_{13}$ the atmospheric angle is fixed to its maximal value, as a consequence of the fact that the model has the 
TBM limit built-in (see next section). Increasing $\theta_{13}$  two different branches above and below $\theta_{23}=\pi/4$ develop and for $\theta_{13}\gsim 1^\circ$  
maximal $2-3$ mixing is strongly excluded. It is interesting to observe that this result can be easily verified at the forthcoming neutrino experiments \cite{Bandyopadhyay:2007kx} 
even with a reduced sensitivity to $\theta_{13}$ since the largest deviation from  $\theta_{23}=\pi/4$ is obtained for $\theta_{13}$ close to its current upper bound. 
We checked that the previous correlation is obtained with $a,b,c$ and $d$ all at the same order of magnitude ${\cal O}(10^{-2})$, which ensures that no fine-tuning 
is required among the matrix elements of eq.(\ref{Eq:mass2}), and with the phase $\phi$ uniformly distributed in the $[-\pi,\pi)$ interval.
 We also verified that the model does not produce any other relevant correlations among the mixing parameters.

\begin{figure}[h!] 
\begin{center} 
\includegraphics[width=10cm]{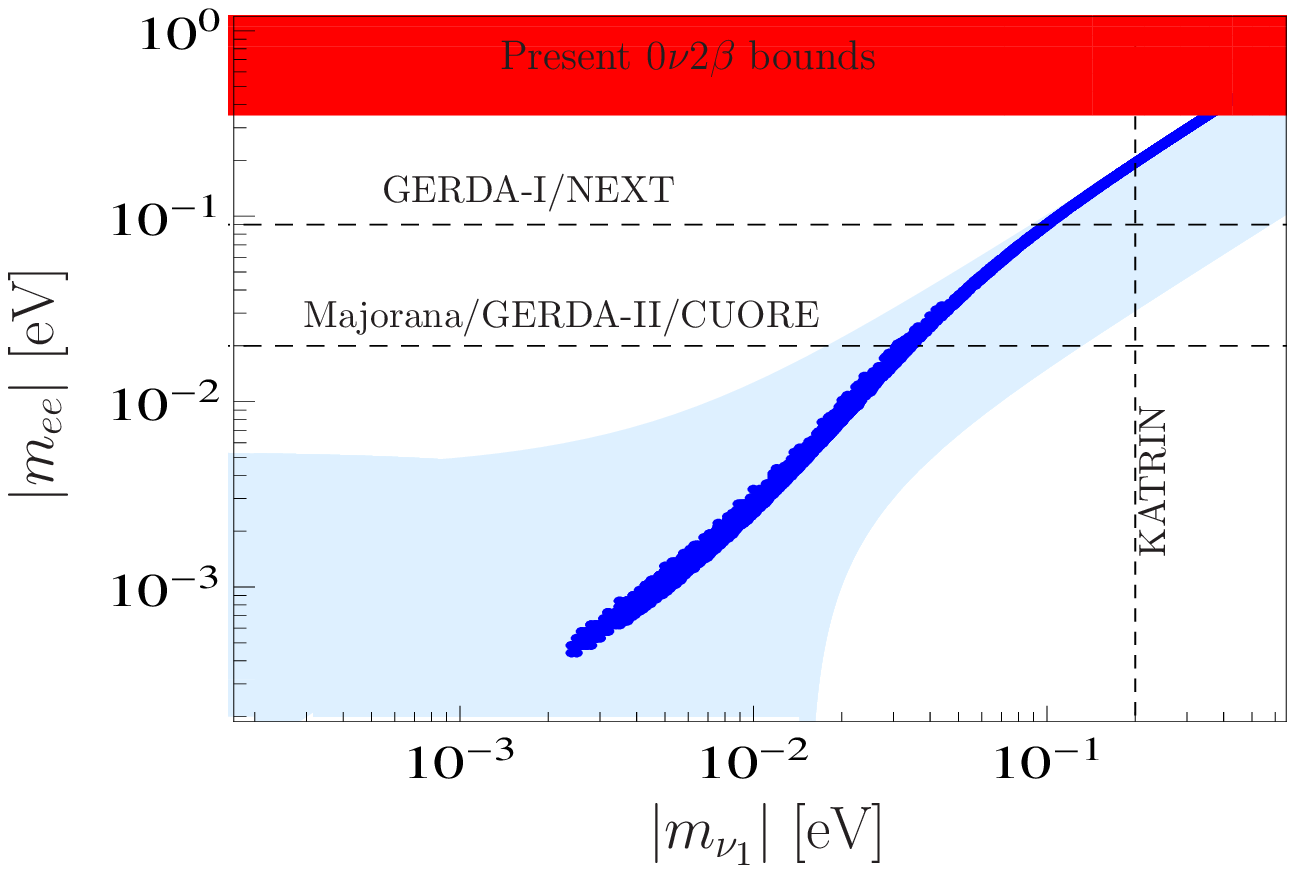}
\caption{\it Values of the effective mass $|m_{ee}|$ (in eV) allowed in our model. The two dashed horizontal
lines represent the experimental sensitivities of some of the forthcoming experiments while
the dashed vertical line is the upper limit from tritium $\beta$-decay experiment. For references to experiments see 
\cite{Osipowicz:2001sq,:2008qu,Smolnikov:2008fu,Giuliani:2008zz,GomezCadenas:2010gs}.}
\label{nu00}
\end{center}
\end{figure}

The other interesting neutrino observable that we want to discuss is the effective mass $|m_{ee}|$ entering the neutrinoless double beta decay rate.
 Since our model is only compatible with the normal hierarchy, we expect it to be quite small for small $m_{\nu_1}$. The result of our simulation can be found in fig.(\ref{nu00}), where we can see that a lower bound 
$(|m_{\nu_1}|, |m_{ee}|)\gsim (2\cdot 10^{-3},4\cdot 10^{-4}) \,eV$  can be set.  The existence of such a lower bound on $|m_{ee}|$ can be justified in the following way: since we are working in the 
basis where the charged leptons are diagonal, a vanishing $|m_{ee}|$ would imply a vanishing parameter $a$; in that case, the mass matrix in eq.(\ref{Eq:mass22}) would have a zero mass
eigenvalue associated to the eigenvector $(1,0,0)^T$, which is not compatible with the neutrino experimental data.

\section{Conclusion}
\label{conc}
In this paper we have studied a model explaining the stability of dark and giving an interesting neutrino phenomenology based on the $A_4$ flavor symmetry. 
The model is an extension of the standard model and contains four Higgs doublets and five
heavy right-handed neutrinos. Light neutrino masses are generated with the type-I seesaw and the resulting Majorana mass matrix has only five free parameters. 
We made a numerical scan of the allowed parameter space and found a correlation between the atmospheric and 
the reactor angles. In particular, for  $\theta_{13}\gsim 1^\circ$ maximal atmospheric angle is strongly disfavoured
whereas for small reactor angle the $\theta_{23}$  is close to maximal. The model gives only normal neutrino 
mass hierarchy and predicts a lower bound for the neutrinoless double beta decay $|m_{ee}|\gsim 4 \cdot 10^{-4}$ eV.

\section{Acknowledgments}
We thank Martin Hirsch for useful discussions.Work of E.P. and S.M.  was supported by the Spanish MICINN under grants
FPA2008-00319/FPA and MULTIDARK CSD2009-00064, by
Prometeo/2009/091, by the EU grant UNILHC PITN-GA-2009-237920.
S.M. was supported by a Juan de la Cierva contract. E.P. was supported by CONACyT (Mexico).
D.M. was supported by the Deutsche Forschungs-gemeinschaft, contract WI 2639/2-1. 
E.P. also acknowledges the Physics Department of the W{\"u}rzburg University  for their hospitality during the 
earlier stages of this work.

\appendix

\section{Mass hierarchy}
\subsection{Neutrino mass hierarchies for the real mass matrix}
\label{real} 
In this section we show that our model allows only the normal hierarchy 
in the simplified case the mass matrix in eq.(\ref{Eq:mass2}) is real. Since the absolute neutrino mass scale is unknown, we can restrict ourselves to study the situation where the lightest neutrino mass eigenvalue is (almost) vanishing. For our discussion it is then enough to consider the determinant of such a matrix, which results:
\bea
{\rm det} (m_\nu) = -a^2 (-b c + d)^2\,.
\eea
For an almost vanishing determinant, we can have two options, namely $a\sim 0$ and $d \sim b\,c$. In the first case, the neutrino mass matrix 
allows a diagonalizing matrix of the form:
\begin{eqnarray}
\label{Eq:mass3}
 m_{\nu} \simeq \left(
  \begin{array}{ccc}
  1 & 0 & 0\\
  0  & x & z \\
  0  & y   & f
  \end{array}
  \right), 
\end{eqnarray}
whose first line implies a vanishing solar mixing angle. The phenomelogical interesting case corresponds then to 
$a\ne 0$ but, even in this case, the null eigenvalue will always be associated with an eigenvector different from $(0,1,-1)^T$,
so it cannot be $m_{\nu_3}$ anyway. This excludes the inverted hierarchy as a viable neutrino mass spectrum.
The second choice $d=b\,c$ always produces two vanishing neutrino masses; if one of them would be $m_{\nu_3}$ then 
the atmospheric mass difference would be vanishing; on the other hand, if $m_{\nu_1} = 0$, then one can have $\Delta m^2_{21} = 0$
but $\Delta m^2_{31} \ne 0$. Then the normal mass scheme provides the only framework where to account for such results.

\subsection{Neutrino mass hierarchies for the vanishing reactor angle}
\label{mutau} 
We work here in the limit of vanishing $\theta_{13}$,  which is a good approximation as given by the 
experimental data. In our case this limit implies $b=c$ in eq. (\ref{Eq:mass22}). This matrix has an eigenvector $(0,-1,1)$ with eigenvalue
\be\label{mtm1}
m_{\nu 3}=(c^2-d e^{i \phi_d}).
\ee
The absolute value of this mass squared is 
\be\label{mtm2}
|m_{\nu 3}|^2=c^4-2 c^2 d \cos \phi_d+d^2.
\ee
From the invariant equations (\ref{invariants}), the determinant of the squared mass matrix is
\be\label{mtdet}
{\rm det}\,(M_\nu^2)= a^4 \left(c^4-2 c^2 d \cos \phi_d+d^2\right)^2 =|m_{\nu_1}|^2\,|m_{\nu_2}|^2|m_{\nu_3}|^2\,.
\ee
From eqs. (\ref{mtm2}) and (\ref{mtdet}) we have the relation
\be\label{mtmbb}
a^4=\frac{|m_{\nu_1}|^2\,|m_{\nu_2}|^2}{|m_{\nu_3}|^2}\,;
\ee
we also know that $|m_{bb}|=a^2$ and therefore
\be\label{mtmbb1}
|m_{bb}|^2=|m_{\nu 1}|^2\frac{|m_{\nu_2}|^2}{|m_{\nu_3}|^2}=|m_{\nu 1}|^2\left(c_{\odot}^4+s_{\odot}^4+s_{\odot}^4\frac{\Delta m^2_{12}}{|m_{\nu 1}|^2}+c_{\odot}^2 s_{\odot}^2 \frac{\sqrt{\Delta m^2_{12}+|m_{\nu 1}|^2}}{|m_{\nu 1}|} \cos \alpha\right)
\ee
where $\alpha$ is the Majorana phase and $c_{\odot}\equiv \cos \theta_{12}$ and $s_{\odot}\equiv \sin \theta_{12}$.
From the previous relation we get: 
\be\label{mtratio}\ba{lcl}\frac{|m_{\nu_2}|^2}{|m_{\nu_3}|^2}&=&\left(c_{\odot}^4+s_{\odot}^4+s_{\odot}^4\frac{\Delta m^2_{12}}{|m_{\nu 1}|^2}+c_{\odot}^2 s_{\odot}^2 \frac{\sqrt{\Delta m^2_{12}+|m_{\nu 1}|^2}}{|m_{\nu 1}|} \cos \alpha\right)\ea\ee
which, for $|m_{\nu 1}|> >\sqrt{\Delta m^2_{12}}$, can be approximated by:
\be\ba{lcl}
\frac{|m_{\nu_2}|^2}{|m_{\nu_3}|^2}&\sim& \left(c_{\odot}^4+s_{\odot}^4+c_{\odot}^2 s_{\odot}^2  \cos \alpha\right)=1-c_{\odot}^2 s_{\odot}^2 (2- \cos \alpha)<1\ea\ee
 
That means that $|m_{\nu 2}|/|m_{\nu 3}|<1$ and implies a normal hierarchy. 
In the limit $m_{\nu 1}\sim \Delta m^2_{12}$ we cannot have the inverse hierarchy because the minimal value for 
$m_{\nu 1}$ is $m^{min}_{\nu 1}=\sqrt{\Delta m^2_{13}}>\sqrt{\Delta m^2_{12}}$.

Notice that the relation (\ref{mtratio}) can be expressed as an implicit function of $m_{\nu 1}$, $\Delta m^2_{12}$, $\Delta m^2_{13}$, $\theta_{12}$ and the Majorana phase $\alpha$. Using the experimental information on the mass differences and the solar angle we can numerically evaluate the minimum allowed value for $m_{\nu 1}$.


\subsection{TBM limit}
\label{app} 
In this appendix we briefly discuss the TBM limit of the mass matrix in eq.(\ref{Eq:mass2}). 
This can be achieved imposing the following relations among the parameters:
\bea 
b&=&c \nn \\
&& \label{tuning} \\
k&=&a^2 + a b -2 b^2 \nn 
\eea
(the first relation is enough to get a $\mu-\tau$ invariant mass matrix).
Then the  masses depend on two complex parameters that can be ordered according to:
\bea
m_1 &=& a (a-b) \nn \\
m_2 &=& a (a+2 b) \nn \\
m_3 &=& -a^2 - a b + 2 b^2 \,.
\eea
It is easier to study the phenomenology redifining
\bea
a &=& |a| \,e^{i \, \phi_a} \nn \\
b &=& |b| \,e^{i \, \phi_b}\nn \\
\frac{|b|}{|a|} &=& t \qquad
\left(\phi_a - \phi_b\right) = \Delta \phi \nn 
\eea 
so that the solar and atmospheric mass differences are:
\bea
\Delta m^2_{21} &=& 3 |a|^4 \,t \,(2 \cos \Delta \phi + t) \nn \\ \label{massdiff} && \\
\Delta m^2_{31} &=& 4  |a|^4 \,t \,( \cos \Delta \phi + t - 2 t \cos^2 \Delta \phi - t^2 \cos \Delta \phi + t^3)\,. \nn
\eea
The model is only compatible with a normal hierarchy spectrum because the simultaneous 
requirements $\Delta m^2_{21}>0$ and $\Delta m^2_{31}<0$ gives
\bea
 -\frac{t}{2}<&\cos \Delta \phi&< 1 \nn \\
&& \\
\cos \Delta \phi&<& -t  \nn
\eea
which are obviously incompatible.
Moreover, it  easy to check that the conditions $|m_1|>0$ and $|m_3|\lesssim 0.5$ eV imply:
\bea
\label{ranget}
0.07 \lesssim t \lesssim 5 \,.
\eea
For $|m_{ee}|$ we found a lower bound $|m_{ee}| > 6 \cdot 10^{-4}$ eV; 
other neutrino mass matrix models with  two complex parameters predicts different lower limits in the TBM limit, for instance  
$|m_{ee}| > 7 \cdot 10^{-3}$eV in \cite{Altarelli:2009kr} and $|m_{ee}|=0$ in \cite{Hirsch:2008rp}.

\end{document}